\documentclass[letterpaper, 10 pt, conference]{ieeeconf}  
\IEEEoverridecommandlockouts                              
\usepackage{comment}

\usepackage{appendix}

\usepackage{etoolbox,mathtools,amsmath,amssymb,amsfonts}
\newcommand{\N}{\mathbb{N}}		
\newcommand{\R}{\mathbb{R}}		
\newcommand{\Rn}{\mathbb{R}^n}		
\newcommand{\Rpl}{\mathbb{R}_+}		
\newcommand{\B}{\mathcal{B}}		
\renewcommand{\P}{\mathcal{P}}		
\DeclareMathOperator*{\conv}{conv}	
\DeclareMathOperator*{\supp}{supp}	
\DeclareMathOperator*{\card}{card}	
\DeclareMathOperator{\lip}{Lip}		

\newcommand{\TV}{\textup{TV}}		
\newcommand{\BL}{\textup{BL}}		

\DeclareMathOperator*{\sign}{sign}		

\usepackage{amsthm}
\theoremstyle{plain}

\theoremstyle{definition}
\newtheorem{definition}{Definition}
\newtheorem{example}{Example}
\newtheorem{assumption}{Assumption}
\theoremstyle{remark}
\newtheorem{remark}{Remark}

\usepackage[createShortEnv]{proof-at-the-end}

\makeatletter
\let\NAT@parse\undefined
\makeatother
\usepackage[numbers,sort&compress]{natbib}
\renewcommand*\citealt[2][]{\citet[#1]{#2}}
\renewcommand*\citealp[2][]{\citep[#1]{#2}}

\usepackage{hyperref}			
\hypersetup{%
	colorlinks=true,%
	linkcolor=black,%
	urlcolor=black,%
	citecolor=black,%
	filecolor=black%
}					
\usepackage[all]{hypcap}		

\usepackage{subcaption}			
\usepackage{booktabs,multirow}		
\usepackage{pgf}
\usepackage{graphicx,pgfplots,tikzscale}	
\usetikzlibrary{calc}			
\pgfplotsset{compat=newest}
\pgfplotsset{plot coordinates/math parser=false}

\newcommand\hl[1]{{#1}}	

\newcommand\proofint{\int}
\newcommand\proofsum{\sum}
\newcommand\prooffrac{\frac}

\title{\LARGE \bf
Choice Paralysis in Evolutionary Games
}

\author{%
	Brendon G.\ Anderson\authorrefmark{1}%
	\thanks{\authorrefmark{1}Mechanical Engineering Department, California Polytechnic State University. Email: {\tt \href{mailto:bga@calpoly.edu}{bga@calpoly.edu}}.}%
}

\begin{document}

\maketitle
\thispagestyle{empty}
\pagestyle{empty}

\begin{abstract}
	In this paper, we consider finite-strategy approximations of infinite-strategy evolutionary games. We prove that such approximations converge to the true dynamics over finite-time intervals, under mild regularity conditions which are satisfied by classical examples, e.g., the replicator dynamics. We identify and formalize novel characteristics in evolutionary games: \emph{choice mobility}, and its complement \emph{choice paralysis}. Choice mobility is shown to be a key sufficient condition for the long-time limiting behavior of finite-strategy approximations to coincide with that of the true infinite-strategy game. An illustrative example is constructed to showcase how choice paralysis may lead to the infinite-strategy game getting ``stuck,'' even though every finite approximation converges to equilibrium.
\end{abstract}

\section{Introduction}
\label{sec: introduction}

Evolutionary games are models in which a (typically large) number of players strategically interact and revise their strategies, in order to maximize their own expected payoffs. Such models arise in a variety of applications ranging from traffic congestion to economic markets \citep{sandholm2010population}. Extensive rigorous analyses of these models have been developed in the case that the players have access to finitely many strategies, with a particular focus on (often control-theoretic) stability guarantees ensuring that the population of players converges to a (Nash) equilibrium state \citep{hofbauer2009stable,fox2013population,mabrok2021passivity,arcak2021dissipativity}. However, many realistic games have a continuum of choices available to players, e.g., in reinforcement learning \citep{mazumdar2020policy}, optimization \citep{anderson2024approximately}, power system pricing and generation \citep{park2001continuous}, and games of timing \citep{bishop1978generalized}. The study of such infinite-strategy evolutionary games becomes significantly more challenging, as the dynamics amount to an infinite-dimensional differential equation in the space of probability measures \citep{anderson2025dissipativity-long}.

A handful of works have analyzed
infinite-strategy models with specific dynamics---most commonly the replicator dynamics \citep{bomze1990dynamical,bomze1991cross,oechssler2001evolutionary,oechssler2002dynamic,cressman2005stability,hingu2018evolutionary,hingu2020superiority}, but also the Brown-von Neumann-Nash, pairwise comparison, logit, and perturbed best response dynamics \citep{hofbauer2009brown,cheung2014pairwise,lahkar2015logit,lahkar2022generalized}, among others. However, even such special cases of infinite-strategy evolutionary games cannot be explicitly solved or numerically simulated, requiring the use of finite-strategy approximations as a tractable surrogate. Unfortunately, such approximations are in general unreliable, as the finite-dimensional dynamics may exhibit asymptotic stability, even though the true infinite-strategy dynamics are unstable \citep{oechssler2002dynamic,anderson2025dissipativity-long}.

In this paper, we study when finite-strategy approximations of infinite-strategy evolutionary dynamics can and cannot be trusted, both over finite-time intervals and in the long-time limit (e.g., convergence to equilibria). A few works have considered finite-strategy approximations for computing Nash equilibria of \emph{static} infinite-strategy games \citep{ganzfried2021algorithm,reny2011strategic}. Though, to the best of our knowledge, the only work to consider the \emph{dynamics} of finite approximations of evolutionary games is \citet{oechssler2002dynamic}, who show that approximations of the replicator dynamics with compact real number interval strategy sets weakly converge to the true dynamics over finite-time intervals. Our work studies the problem for more general dynamics and strategy sets---recovering \citet[Theorem~4]{oechssler2002dynamic} as a special case of our \autoref{thm: approx}---and we provide novel characterizations of approximations in the long-time limit. Specifically, our primary contributions are as follows.

\subsection{Contributions}
\begin{enumerate}
	\item We prove that, under mild regularity conditions, the trajectories of finite-dimensional approximations weakly converge to the true infinite-dimensional evolutionary dynamics on compact time intervals.
	\item We introduce the key notion of \emph{choice mobility}, and its complement \emph{choice paralysis}.
	\item We prove that choice mobility is sufficient to ensure coincidence between the long-time limiting behavior of finite approximations and the infinite-dimensional dynamics (again under mild regularity assumptions).
	\item We show that choice paralysis arises if strategy switching rates decrease as more strategies are introduced.
	\item We construct an example demonstrating that choice paralysis may manifest in convergence of every finite approximation to equilibrium, despite the infinite-strategy game being stuck at a different state.
\end{enumerate}

\hl{Our results imply that numerically tractable finite-strategy approximations of sufficiently fine resolution can be trusted to accurately simulate both the short- and long-time dynamics of infinite-strategy games. They also suggest that the long-time behavior of such approximations \emph{cannot} be trusted in cases where choice paralysis arises, motivating the need for the future development of new analysis techniques.}
Proofs are deferred to the appendix.

\section{Infinite-Strategy Evolutionary Games}
\label{sec: prelims}


We consider a \hl{population} game in which a \hl{continuum} of players choose strategies from an infinite compact metric space $S$ with metric $d_S$. \hl{This level of generality for $S$ allows it to represent either pure or mixed strategies.} The sets of Borel probability measures and continuous real-valued functions on $S$ are denoted by $\P(S)$ and $C(S)$, respectively. The Borel $\sigma$-algebra on $S$ is $\B(S)$. \hl{The Dirac measure at $s\in S$ is denoted by $\delta_s \in \P(S)$.} The \emph{state} of the population is encoded by a probability measure, $\mu \in \P(S)$, over the strategy set $S$, representing the frequency with which each strategy is played. \hl{The \emph{game}, $F \colon \P(S) \to C(S)$, is defined by \emph{average payoffs} $(F(\mu))(s)$ that encode how much payoff a player choosing strategy $s\in S$ receives when the population as a whole is playing strategies according to the distribution $\mu$. \hl{As is standard in the study of population games, the population is assumed to be homogeneous, with all players sharing the same strategy set $S$ and game $F$---see \citet[Section~8.3.2]{sandholm2010population} for discussion on heterogeneous multipopulation extensions.} Every player aims to choose a strategy $s$ that maximizes their payoff $(F(\mu))(s)$.}

The population state evolves according to an evolutionary dynamics model (EDM) \citep{park2019population,anderson2025dissipativity-long} taking the form
\begin{equation}
	\dot{\mu}(t) = v(\mu(t),\rho(t)), \quad \rho(t) = F(\mu(t)),
	\label{eq: edm}
\end{equation}
where $v \colon \P(S) \times C(S) \to T\P(S)$, with $T\P(S)$ being the tangent space of $\P(S)$, defines the dynamics, and $\rho$ encodes the game's payoffs that steer the dynamics. The EDM \eqref{eq: edm} can be viewed as a state feedback control system with open-loop dynamics $v$ and controller $F$. \hl{As such, the design of incentive and disincentive structures, encoded into $F$, is the population-level analogue of traditional feedback control design tasks, with applications to societal-scale systems (e.g., traffic networks \citealp{sandholm2010population}).} The time-derivative $\dot{\mu}(t)$ is defined in the Fr\'echet sense. \hl{Under the standard assumption of $F$ being weakly continuous, the game admits a Nash equilibrium \citep{glicksberg1952further}, at which standard EDMs---such as those we consider---come to rest. Such equilibria need not be asymptotically stable \citep{anderson2025dissipativity-long}.} Throughout, we assume that \eqref{eq: edm} admits a unique and weakly continuous solution $\mu \colon [0,\infty) \to \P(S)$.

We consider a general class of dynamics, termed \emph{mean dynamics} \citep{sandholm2010population,cheung2014pairwise}, where the dynamics map $v$ takes the form
\begin{equation}
	\begin{aligned}
		v(\mu,\rho)(B) &= \int_S \int_B \varphi(s,s',\mu,\rho) d\lambda(s') d\mu(s) \\
		& \qquad - \int_S \int_B \varphi(s',s,\mu,\rho) d\mu(s') d\lambda(s)
	\end{aligned}
	\label{eq: mean_dynamics_map}
\end{equation}
for some reference probability measure $\lambda\in\P(S)$ and some adequately integrable map $\varphi \colon S\times S \times \P(S) \times C(S) \to \Rpl$, with $\Rpl$ denoting the set of nonnegative real numbers.
\hl{The map $\varphi$ is known as the \emph{revision protocol}, as it characterizes the rate $\varphi(s,s',\mu,\rho)$ at which players switch from one strategy $s$ to another $s'$, given a particular state $\mu$ and payoff profile $\rho$ of the game}. The first term in $v(\mu,\rho)(B)$ represents the ``inflow'' of mass into the strategies within $B\in\B(S)$, and the second term represents the corresponding ``outflow.''

\begin{example}
	\label{ex: revision_protocols}
	Popular examples of mean dynamics include the replicator dynamics \citep{bomze1990dynamical,bomze1991cross,oechssler2001evolutionary,oechssler2002dynamic}, the Brown-von Neumann-Nash (BNN) dynamics \citep{hofbauer2009brown}, and the pairwise comparison dynamics \citep{cheung2014pairwise}. Specifically, the replicator dynamics correspond to setting $\lambda = \mu$ and $\varphi(s,s',\mu,\rho) = \max\{0,\rho(s') - \rho(s)\}$. The BNN and pairwise comparison dynamics correspond to a fixed-in-time $\lambda$ (typically uniform), and are respectively defined by $\varphi(s,s',\mu,\rho) = \max\left\{0,\rho(s') - \int_S \rho d\mu\right\}$ and $\varphi(s,s',\mu,\rho) = \gamma(s,s',\rho)$ for some function $\gamma$ independent of $\mu$, assumed to satisfy \emph{sign-preservation}: the switch rate is positive if and only if the new strategy has higher payoff than the old strategy, i.e., $\sign(\gamma(s,s',\rho)) = \sign(\max\{0,\rho(s') - \rho(s)\})$ for all $s,s'\in S$ and all $\rho\in C(S)$. The pairwise comparison dynamics are called \emph{impartial} if, for all $s'\in S$, there exists a continuous function $\phi_{s'} \colon \R \to \Rpl$ satisfying $\gamma(s,s',\rho) = \phi_{s'}(\rho(s') - \rho(s))$ for all $s\in S$ and all $\rho \in C(S)$.
\end{example}

The EDM \eqref{eq: edm} is an infinite-dimensional differential equation evolving in the space of probability measures, and hence cannot be tractably solved in general. This motivates the problem studied in this paper: introduce and analyze finite approximations, and characterize when such approximations accurately reflect the true behavior of the evolutionary game.

\section{Finite-Strategy Approximations}
\label{sec: approx}

\subsection{Formulation}
\label{sec: approx_formulation}

We now explicitly formulate finite-dimensional approximations of the EDM \eqref{eq: edm} with the mean dynamics map \eqref{eq: mean_dynamics_map}. Consider $n\in\N$, and let $\lambda_n,\mu_n(0)\in\P(S)$ be discrete measures with respect to some reference measure $\nu$ on $S$ (e.g., Lebesgue measure when $S$ is a subset of Euclidean space). Suppose that $\supp(\lambda_n) = \supp(\mu_n(0)) = S_n \coloneqq \{s_1,\dots,s_n\} \subseteq S$, so that $\lambda_n = \sum_{i=1}^n \lambda_n(s_i) \delta_{s_i}$ and $\mu_n(0) = \sum_{i=1}^n (\mu_n(0))(s_i) \delta_{s_i}$. Here, we write $\lambda_n(s_i)$ to mean $\lambda_n(\{s_i\})$ and similarly for other measures evaluated on singleton sets. Let $\mu_n\colon [0,\infty) \to \P(S)$ denote the solution to the EDM with reference measure $\lambda_n$ and initial state $\mu_n(0)$. The measures $\lambda_n$ and $\mu_n(t)$ respectively serve as approximations to $\lambda$ and $\mu(t)$, which may have full support. Writing the mean dynamics for the approximation yields
\begin{align*}
	&(\dot{\mu}_n(t))(B) \\
	&\quad = \int_S\int_B \varphi(s,s',\mu_n(t),F(\mu_n(t))) d\lambda_n(s') d(\mu_n(t))(s) \\
	&\quad\quad - \int_S\int_B \varphi(s',s,\mu_n(t),F(\mu_n(t))) d(\mu_n(t))(s') d\lambda_n(s)
\end{align*}
for all $B\in\B(S)$. Since $\lambda_n(B) = (\mu_n(0))(B) = 0$ for $B = S\setminus S_n$, we immediately see that $(\dot{\mu}_n(0))(B) = 0$ and hence $(\mu_n(t))(B) = 0$ for all $t\in[0,\infty)$ for this set $B$. Thus, $\mu_n(t)$ is discrete with $\supp(\mu_n(t)) \subseteq S_n$ for all $t \in [0,\infty)$, implying that it takes the form
\begin{equation}
	\mu_n(t) = \sum_{i=1}^n (\mu_n(t))(s_i) \delta_{s_i}.
	\label{eq: finite_measure}
\end{equation}
Thus, we may uniquely identify $\mu_n$ with the curve $x_n \colon [0,\infty) \to \Delta^{n-1}$ defined by $(x_n(t))_i = (\mu_n(t))(s_i)$, where $\Delta^{n-1}$ denotes the probability simplex in $\Rn$.

We define a finite-dimensional game $\hat{F}_n\colon \Delta^{n-1} \to \Rn$ by restricting $F$ to discrete measures on $S_n$:
\begin{equation*}
	(\hat{F}_n(\theta))_i = F\Bigg(\sum_{j=1}^n \theta_j \delta_{s_j}\Bigg)(s_i), ~ i\in\{1,\dots,n\}.
\end{equation*}
Similarly, we define a finite-dimensional revision protocol $\hat{\varphi}_n \colon \{1,\dots,n\}\times \{1,\dots,n\} \times \Delta^{n-1} \to \Rpl$ by
\begin{equation*}
	\hat{\varphi}_n(i,j,\theta) = \varphi\Bigg(s_i,s_j,\sum_{k=1}^n \theta_k \delta_{s_k}, F\Bigg(\sum_{k=1}^n \theta_k \delta_{s_k}\Bigg)\Bigg).
\end{equation*}
Under these constructions, the mean dynamics of the finite approximation are equivalently rewritten as
\begin{equation}
	\begin{aligned}
		(\dot{x}_n(t))_i &= \lambda_n(s_i) \sum_{j=1}^n \hat{\varphi}_n(j,i,x_n(t)) (x_n(t))_j \\
		&\quad - (x_n(t))_i \sum_{j=1}^n \hat{\varphi}_n(i,j,x_n(t))\lambda_n(s_j),
	\end{aligned}
	\label{eq: n_ode}
\end{equation}
for all $i\in\{1,\dots,n\}$. Note that the approximation \eqref{eq: n_ode} is a system of $n$ ordinary differential equations, and hence can be solved efficiently using off-the-shelf numerical solvers.

\begin{example}
	\label{ex: finite_approx}
	For the replicator dynamics of \autoref{ex: revision_protocols}, the above constructions give $\lambda_n(s_i) = (x_n(t))_i$ and $\hat{\varphi}_n(i,j,\theta) = \max\left\{0, (\hat{F}_n(\theta))_j - (\hat{F}_n(\theta))_i \right\}$, and therefore \eqref{eq: n_ode} reduces to
	\begin{equation*}
		(\dot{x}_n(t))_i = (x_n(t))_i \left((\hat{F}_n(x_n(t)))_i - x_n(t)^\top \hat{F}_n(x_n(t))\right),
	\end{equation*}
	for all $i\in\{1,\dots,n\}$, which coincides with the classical replicator dynamics on the finite strategy set $\{1,\dots,n\}$ (see \citealt[Example~4.3.1]{sandholm2010population}). Similar recovery of the classical finite-dimensional BNN dynamics \citep[Example~4.3.4]{sandholm2010population} and pairwise comparison dynamics \citep[Example~4.3.5]{sandholm2010population} are also achieved (specifically, under the fixed uniform reference measure given by $\lambda_n(s_i) = \frac{1}{n}$ for all $i$).
\end{example}

\subsection{Finite-Time Analysis}
In this section, we show that, under mild regularity conditions on $\varphi$, approximations to the EDM \eqref{eq: edm} weakly converge to the true dynamics on compact time intervals. Throughout our analyses, we denote the bounded-Lipschitz metric on $\P(S)$ by $d_\BL \colon (\nu,\eta) \mapsto \sup\left\{ \left| \int_S g d\nu - \int_S g d\eta \right| : g\in \lip_1(S), ~ \|g\|_\infty \le 1 \right\}$, where $\lip_L(S)$ denotes the set of all $L$-Lipschitz real-valued functions on $S$ and $\|\cdot\|_\infty$ denotes the supremum norm. Recall that $d_\BL$ metrizes the weak topology on $\P(S)$ \citep[Theorem~8.3.2]{bogachev2007measure}. We now formally state the regularity conditions:

\begin{assumption}
	\label{ass: approx}
	The following all hold:
	\begin{enumerate}
		\item {\bf Bounded.} There exists $M\in\Rpl$ such that $|\varphi(s,s',\mu,F(\mu))| \le M$ for all $s,s'\in S$, $\mu\in\P(S)$.
		\item {\bf Lipschitz on $S$.} There exist $L_1,L_2\in\Rpl$ such that
		\begin{gather*}
		\resizebox{\linewidth}{!}{%
			$\varphi(\cdot,s',\mu,F(\mu)) \in \lip_{L_1}(S) ~ \text{for all $s'\in S, ~ \mu \in \P(S)$},$%
		} \\
			\varphi(s,\cdot,\mu,F(\mu)) \in \lip_{L_2}(S) ~ \text{for all $s\in S, ~ \mu \in \P(S)$}.
		\end{gather*}
		\item {\bf Lipschitz on $\P(S)$.} There exists $L_3\in\Rpl$ such that, for all $s,s'\in S$, $\mu,\mu' \in \P(S)$, it holds that
		\begin{equation*}
		\resizebox{\linewidth}{!}{%
			$|\varphi(s,s',\mu,F(\mu)) - \varphi(s,s',\mu',F(\mu'))| \le L_3 d_\BL(\mu,\mu').$%
		}
		\end{equation*}
	\end{enumerate}
\end{assumption}

Notice that all conditions in \autoref{ass: approx} are with respect to the closed-loop dynamics, i.e., they are required to hold for state-payoff pairs $(\mu,\rho)$ satisfying $\rho = F(\mu)$. In general, this is much less stringent than requiring them to hold for all states $\mu\in\P(S)$ and all payoff functions $\rho\in C(S)$.

The replicator, BNN, and impartial pairwise comparison dynamics (recall \autoref{ex: revision_protocols}) all satisfy \autoref{ass: approx} for typical games, and thus the conditions are considered mild:

\begin{propositionE}[][end,text link=]
	\label{prop: bnn_pcd_approx}
	Consider either the replicator dynamics, the BNN dynamics, or the impartial pairwise comparison dynamics with $(s',p) \mapsto \phi_{s'}(p)$ Lipschitz with respect to the metric $d_{S\times \R} \colon ((s',p),(\tilde{s}',\tilde{p})) \mapsto \| (d_S(s',\tilde{s}'), |p-p'|) \|$ where $\|\cdot\|$ denotes an arbitrary norm on $\R^2$ (e.g., the Smith dynamics \citep{cheung2014pairwise}). If $F$ is Lipschitz continuous between $(\P(S),d_\BL)$ and $(C(S),\|\cdot\|_\infty)$ and there exists $L\in\Rpl$ such that $F(\mu)\in\lip_L(S)$ for all $\mu\in\P(S)$, then \autoref{ass: approx} holds. In particular, this is the case for the commonly considered linear games $F$ taking the form
	\begin{equation*}
		F(\mu)(s) = \int_S f(s,s')d\mu(s')
	\end{equation*}
	with $f \colon S\times S \to \R$ bounded and Lipschitz with respect to the metric $d_{S\times S} \colon ((s,s'),(\tilde{s},\tilde{s}')) \mapsto \|(d_S(s,\tilde{s}),d_S(s',\tilde{s}'))\|$.
\end{propositionE}

\begin{proofE}
	We first prove that the hypotheses on $F$ hold when $F$ is defined by
	\begin{equation}
		F(\mu)(s) = \proofint_S f(s,s')d\mu(s'),
		\label{eq: linear_payoff}
	\end{equation}
	with $f\colon S\times S \to \R$ bounded and Lipschitz with respect to $d_{S\times S}$. Let $M_f\in\Rpl$ and $L_f\in\Rpl$ be the associated boundedness and Lipschitzness constants of $f$, so that
	\begin{equation*}
		\sup_{(s,s')\in S\times S} |f(s,s')| \le M_f,
	\end{equation*}
	and
	\begin{equation*}
		|f(s,s') - f(\tilde{s},\tilde{s}')| \le L_f d_{S\times S}((s,s'),(\tilde{s},\tilde{s}'))
	\end{equation*}
	for all $(s,s'),(\tilde{s},\tilde{s}')\in S\times S$. It is easy to see that, since $d_{S\times S}((s,s'),(\tilde{s},\tilde{s}')) = \|(d_S(s,\tilde{s}),d_S(s',\tilde{s}'))\|_p$ for some $p$-norm on $\R^2$, it holds that $f(s,\cdot)$ and $f(\cdot,s')$ are $L_f$-Lipschitz on $S$ with respect to $d_S$, for all $s,s'\in S$.

	Let $\mu,\mu'\in\P(S)$. Then
	\begin{align*}
		&\|F(\mu) - F(\mu')\|_\infty \\
		&\quad= \sup_{s\in S}\left|\proofint_S f(s,s') d\mu(s') - \proofint_S f(s,s') d\mu'(s')\right| \\
		&\quad= \sup_{s\in S} \left|\proofint_S f(s,s')d(\mu-\mu')(s')\right| \\
		&\quad\le \max\{M_f,L_f\} d_\BL(\mu, \mu'),
	\end{align*}
	which proves that $F$ is Lipschitz continuous with respect to $d_\BL$. Furthermore, we find that, for all $s,s'\in S$, it holds that
	\begin{align*}
		|F(\mu)(s) - F(\mu)(s')| &= \left|\proofint_S (f(s,\tilde{s}) - f(s',\tilde{s}))d\mu(\tilde{s})\right| \\
		&\le \proofint_S |f(s,\tilde{s}) - f(s',\tilde{s})|d|\mu|(\tilde{s}) \\
		&\le L_f d_S(s,s').
	\end{align*}
	This proves that $F(\mu)\in \lip_{L_f}(S)$ for all $\mu\in\P(S)$, and hence that $F$ indeed satisfies the hypotheses of the proposition when given by \eqref{eq: linear_payoff}.

	Now suppose that $F$ is a general map (i.e., it does not necessarily take the form \eqref{eq: linear_payoff}) that is Lipschitz continuous with respect to $d_\BL$, so that there exists $K\in\Rpl$ such that
	\begin{equation*}
		\|F(\mu) - F(\mu')\|_\infty \le K d_\BL(\mu, \mu').
	\end{equation*}
	Furthermore, suppose that there exists $L\in\Rpl$ such that $F(\mu)\in\lip_L(S)$ for all $\mu\in\P(S)$. We will now prove the result for the three dynamics separately, ending with the replicator dynamics.
	
	\setcounter{paragraph}{0}
	\paragraph{BNN dynamics}

	Consider the BNN dynamics. The revision protocol is defined by
	\begin{equation*}
		\varphi(s,s',\mu,\rho) = \max\{0,\left<\rho,\delta_{s'}\right> - \left<\rho,\mu\right>\}.
	\end{equation*}
	
	Let $s,s'\in S$ and let $\mu\in\P(S)$. It holds that
	\begin{align*}
		|\varphi(s,s',\mu,F(\mu))| &\le |\left<F(\mu),\delta_{s'} - \mu\right>| \\
		&\le \|F(\mu)\|_\infty \|\delta_{s'} - \mu\|_\TV \\
		&\le 2\|F(\mu)\|_\infty \\
		&\le M
	\end{align*}
	with $M \coloneqq \frac{1}{2}\sup_{\mu' \in\P(S)}\|F(\mu')\|_\infty$ finite, since $F$ being Lipschitz continuous with respect to $d_\BL$ implies that $F$ is weakly continuous, and hence $\mu' \mapsto \|F(\mu')\|_\infty$ is a weakly continuous map from the weakly compact set $\P(S)$ to $\Rpl$ (where weak compactness of $\P(S)$ comes from \citealt[Theorem~6.4]{parthasarathy1967probability}). This proves the first condition in \autoref{ass: approx}.

	Next, notice that, for all $s'\in S$ and all $\mu\in\P(S)$, the function $s\mapsto \varphi(s,s',\mu,F(\mu))$ is constant and hence Lipschitz with constant $0$. Let $s\in S$ and let $\mu\in\P(S)$. We see that
	\begin{align*}
		s' \mapsto \left<F(\mu),\delta_{s'}\right> - \left<F(\mu),\mu\right> = F(\mu)(s') - \left<F(\mu),\mu\right>
	\end{align*}
	is $L$-Lipschitz on $S$ since $F(\mu)\in \lip_L(S)$. Therefore, the composition
	\begin{equation*}
		s'\mapsto \varphi(s,s',\mu,F(\mu)) = \max\{0,\left<F(\mu),\delta_{s'}\right> - \left<F(\mu),\mu\right>\}
	\end{equation*}
	is also $L$-Lipschitz on $S$. Since $s$ and $\mu$ are arbitrary, this proves the second condition in \autoref{ass: approx}.

	Finally, let $s,s'\in S$ and let $\mu,\mu'\in\P(S)$. It holds that
	\begin{align*}
		& |\left<F(\mu),\delta_{s'}\right> - \left<F(\mu),\mu\right> - \left<F(\mu'),\delta_{s'}\right> + \left<F(\mu'),\mu'\right>| \\
		& \quad \le |\left<F(\mu)-F(\mu'),\delta_{s'}\right>| + |\left<F(\mu),\mu\right> - \left<F(\mu),\mu'\right>| \\
		&\quad\quad+ |\left<F(\mu),\mu'\right> - \left<F(\mu'),\mu'\right>| \\
		&\quad \le \|F(\mu)-F(\mu')\|_\infty + \left|\proofint_S F(\mu) d(\mu-\mu')\right| \\
		&\quad\quad+ \|F(\mu)-F(\mu')\|_\infty \\
		&\quad \le 2K d_\BL(\mu,\mu') + \left|\proofint_S F(\mu) d(\mu-\mu')\right|.
	\end{align*}
	Since $F(\mu) \in \lip_L(S)$ and $\|F(\mu)\|_\infty \le 2M$, it holds that $F(\mu)$ is a bounded Lipschitz function, and we find that
	\begin{equation*}
		\left|\proofint_S F(\mu) d(\mu-\mu')\right| \le \max\{L,2M\} d_\BL(\mu,\mu').
	\end{equation*}
	Thus,
	\begin{align*}
		& |\left<F(\mu),\delta_{s'}\right> - \left<F(\mu),\mu\right> - \left<F(\mu'),\delta_{s'}\right> + \left<F(\mu'),\mu'\right>| \\
		& \quad \le (2K+\max\{L,2M\}) d_\BL(\mu,\mu').
	\end{align*}
	Since $s,s'\in S$ and $\mu,\mu'\in\P(S)$ were chosen arbitrarily, we conclude that
	\begin{equation*}
		\mu \mapsto \left<F(\mu),\delta_{s'}\right> - \left<F(\mu),\mu\right>
	\end{equation*}
	is Lipschitz continuous with respect to $d_\BL$ with constant $2K+\max\{L,2M\}$ for all $s,s'\in S$. Since $x\mapsto \max\{0,x\}$ is a real-valued $1$-Lipschitz function on $\R$, this proves Lipschitzness of the composition, i.e., that
	\begin{equation*}
		\begin{multlined}
		|\varphi(s,s',\mu,F(\mu)) - \varphi(s,s',\mu',F(\mu'))| \\
		\le (2K+\max\{L,2M\}) d_\BL(\mu,\mu')
		\end{multlined}
	\end{equation*}
	for all $s,s'\in S$ and all $\mu,\mu'\in\P(S)$. This concludes the proof for the BNN dynamics.

	\paragraph{Impartial pairwise comparison dynamics}

	Consider the impartial pairwise comparison dynamics. The revision protocol is defined by
	\begin{equation*}
		\varphi(s,s',\mu,\rho) = \phi_{s'}(\rho(s') - \rho(s)),
	\end{equation*}
	where, under our assumptions, $\phi_{s'}$ is a nonnegative $L'$-Lipschitz function on $\R$ satisfying $\phi_{s'}(0) = 0$ by sign-preservation.
	
	Let $s,s'\in S$ and let $\mu\in\P(S)$. It holds that
	\begin{align*}
		|\varphi(s,s',\mu,F(\mu))| &= |\phi_{s'}(F(\mu)(s') - F(\mu)(s))| \\
		&= |\phi_{s'}(F(\mu)(s') - F(\mu)(s)) - \phi_{s'}(0)| \\
		&\le L' |F(\mu)(s') - F(\mu)(s) - 0| \\
		&\le \|F(\mu)\|_\infty \|\delta_{s'} - \delta_s\|_\TV \\
		&\le 2\|F(\mu)\|_\infty \\
		&\le M
	\end{align*}
	with $M \coloneqq \frac{1}{2}\sup_{\mu' \in\P(S)}\|F(\mu')\|_\infty$ finite, since $F$ being Lipschitz continuous with respect to $d_\BL$ implies that $F$ is weakly continuous, and hence $\mu' \mapsto \|F(\mu')\|_\infty$ is a weakly continuous map from the weakly compact set $\P(S)$ to $\Rpl$ (where weak compactness of $\P(S)$ comes from \citealt[Theorem~6.4]{parthasarathy1967probability}). This proves the first condition in \autoref{ass: approx}.

	Next, let $s' \in S$ and let $\mu\in\P(S)$. We see that
	\begin{equation*}
		s\mapsto F(\mu)(s') - F(\mu)(s)
	\end{equation*}
	is $L$-Lipschitz on $S$ since $F(\mu)\in\lip_L(S)$. Therefore, the composition
	\begin{equation*}
		s\mapsto \varphi(s,s',\mu,F(\mu)) = \phi_{s'}(F(\mu)(s) - F(\mu)(s'))
	\end{equation*}
	is Lipschitz continuous on $S$ with constant $L'\cdot L$. Following the same line of analysis shows that, for all $s\in S$ and all $\mu\in\P(S)$, it holds that
	\begin{equation*}
		s' \mapsto \varphi(s,s',\mu,F(\mu))
	\end{equation*}
	is Lipschitz continuous on $S$ with constant $L'\cdot L$. Therefore, the second condition in \autoref{ass: approx} holds.

	Finally, let $s,s'\in S$ and let $\mu,\mu'\in\P(S)$. It holds that
	\begin{align*}
		&|F(\mu)(s') - F(\mu)(s) - F(\mu')(s') + F(\mu')(s)| \\
		&\quad \le |F(\mu)(s') - F(\mu')(s')| + |F(\mu)(s) - F(\mu')(s)| \\
		&\quad \le 2\|F(\mu) - F(\mu')\|_\infty \\
		&\quad \le 2K d_\BL(\mu,\mu').
	\end{align*}
	Since $s,s'\in S$ and $\mu,\mu'\in\P(S)$ were chosen arbitrarily, we conclude that
	\begin{equation*}
		\mu \mapsto F(\mu)(s') - F(\mu)(s)
	\end{equation*}
	is Lipschitz continuous with respect to $d_\BL$ with constant $2K$ for all $s,s'\in S$. Since $\phi_{s'}$ is a real-valued $L'$-Lipschitz function on $\R$, this proves Lipschitzness of the composition, i.e., that
	\begin{equation*}
		|\varphi(s,s',\mu,F(\mu)) - \varphi(s,s',\mu',F(\mu'))| \le 2KL' d_\BL(\mu,\mu')
	\end{equation*}
	for all $s,s'\in S$ and all $\mu,\mu'\in\P(S)$. This concludes the proof for the impartial pairwise comparison dynamics.

	\paragraph{Replicator dynamics}
	Consider the replicator dynamics. The revision protocol is defined by
	\begin{equation*}
		\varphi(s,s',\mu,\rho) = \max\{0, \rho(s') - \rho(s)\}.
	\end{equation*}
	Notice that $\max\{0,\cdot\}$ is a nonnegative $1$-Lipschitz function on $\R$ satisfying $\max\{0,0\} = 0$. Thus, the proof for the replicator dynamics follows immediately from the above proof for the impartial pairwise comparison dynamics with the particular function $\phi_{s'} \coloneqq \max\{0,\cdot\}$.
\end{proofE}

We now state our finite-time approximation result.

\begin{theoremE}[][end,text link=]
	\label{thm: approx}
	Assume \autoref{ass: approx}.
	For all $n\in\N$, consider approximations $\lambda_n,\mu_n(0) \in \P(S)$ to $\lambda$ and $\mu(0)$, with associated solutions $\mu_n$ to the mean dynamics EDM \eqref{eq: edm}--\eqref{eq: mean_dynamics_map}. Assume that $\lambda$ and all $\lambda_n$ are fixed in time. If $\lambda_n \overset{n \to \infty}{\to} \lambda$ weakly and $\mu_n(0) \overset{n\to\infty}{\to} \mu(0)$ weakly, then
	\begin{equation}
		\lim_{n\to \infty} \sup_{t\in[0,T]} d_\BL(\mu_n(t), \mu(t)) = 0 ~ ~  \text{for all} ~ ~ T\in(0,\infty).
		\label{eq: approx}
	\end{equation}
\end{theoremE}

\begin{proofE}
	To simplify the exposition, we denote $\varphi(s,s',\mu(\tau),F(\mu(\tau)))$ by $\varphi(s,s',\tau)$ and $\varphi(s,s',\mu_n(\tau),F(\mu_n(\tau)))$ by $\varphi_n(s,s',\tau)$ for all $\tau \in [0,\infty)$.

	Let $T\in(0,\infty)$. Furthermore, let $n\in\N$, let $t\in[0,T]$, and let $g\in\lip_1(S)$ be such that $\|g\|_\infty \le 1$. It holds that
	\begin{gather*}
		\begin{aligned}
		&\proofint_S g d\mu(t) \\
		&\quad= \proofint_{[0,t]}\proofint_S g(s) d(\dot{\mu}(\tau))(s) d\tau + \proofint_S g(s) d(\mu(0))(s) \\
		&\quad= \proofint_{[0,t]}\proofint_S g(s) \proofint_S \varphi(s',s,\tau) d(\mu(\tau))(s') d\lambda(s) d\tau \\
		&\quad\quad- \proofint_{[0,t]}\proofint_S g(s) \proofint_S \varphi(s,s',\tau) d\lambda(s') d(\mu(\tau))(s) d\tau \\
		& \quad\quad + \proofint_S g(s) d(\mu(0))(s),
		\end{aligned} 
	\end{gather*}
	and an analogous expression holds relating $\mu_n$, $\varphi_n$, and $\lambda_n$.
	Our goal is to estimate $\proofint_S g d\mu(t) - \proofint_S g d\mu_n(t)$ by comparing corresponding ``inflows,'' ``outflows,'' and initial states.

	\setcounter{paragraph}{0}
	\paragraph{Inflows}
	Define
	\begin{align*}
		\Delta_1(\tau) &\coloneqq \proofint_S g(s) \proofint_S \varphi(s',s,\tau) d(\mu(\tau))(s') d\lambda(s) \\
		& \quad - \proofint_S g(s) \proofint_S \varphi_n(s',s,\tau) d(\mu_n(\tau))(s') d\lambda_n(s). 
	\end{align*}
	We have that
	\begin{align*}
		& \left|\proofint_S g(s) \proofint_S (\varphi(s',s,\tau) - \varphi_n(s',s,\tau)) d(\mu(\tau))(s') d\lambda(s) \right| \\
		&\hspace*{0.25em} = \left|\proofint_S g(s) \proofint_S (\varphi(s',s,\tau) - \varphi_n(s',s,\tau)) d(\mu(\tau))(s') d\lambda(s) \right| \\
		& \hspace*{0.25em} \le \proofint_S |g(s)| \proofint_S L_3 d_\BL( \mu(\tau), \mu_n(\tau)) d(\mu(\tau))(s') d\lambda(s) \\
		&\hspace*{0.25em} \le L_3 d_\BL(\mu(\tau), \mu_n(\tau)),
	\end{align*}
	which follows from the bound $\|g\|_\infty \le 1$ and the Lipschitzness of $\varphi$ on $\P(S)$. Next,
	\begin{align*}
		& \left|\proofint_S g(s) \proofint_S \varphi_n(s',s,\tau)d(\mu(\tau) - \mu_n(\tau))(s') d\lambda(s)\right| \\
		&\quad = \left|\proofint_S g(s) \proofint_S \varphi_n(s',s,\tau)d(\mu(\tau) - \mu_n(\tau))(s') d\lambda(s)\right| \\
		&\quad \le \max\{L_1,M\} d_\BL(\mu(\tau), \mu_n(\tau))
	\end{align*}
	follows from boundedness and Lipschitzness of $\varphi$ in its first argument together with the bound $\|g\|_\infty \le 1$. Furthermore,
	\begin{align*}
		& \left|\proofint_S g(s) \proofint_S \varphi_n(s',s,\tau) d(\mu_n(\tau))(s') d(\lambda - \lambda_n)(s)\right| \\
		& \quad =\left|\proofint_S\proofint_S g(s) \varphi_n(s',s,\tau) d(\lambda - \lambda_n)(s) d(\mu_n(\tau))(s')\right| \\
		&\quad \le (L_2+M) d_\BL(\lambda, \lambda_n),
	\end{align*}
	since $s\mapsto g(s)\varphi_n(s',s,\tau)$ is bounded with constant $M$ and is Lipschitz with constant $L_2+M$ (which relies on boundedness of both $g$ and $\varphi$; in general, the product of Lipschitz functions is not Lipschitz). Thus, triangle inequality gives
	\begin{align*}
		|\Delta_1(\tau)| &\le (L_3 + \max\{L_1,M\}) d_\BL(\mu(\tau), \mu_n(\tau)) \\
		&\quad + (L_2+M) d_\BL(\lambda, \lambda_n).
	\end{align*}

	\paragraph{Outflows}
	From a nearly identical analysis as above:
	\begin{align*}
		\Delta_2(\tau) &\coloneqq \proofint_S g(s) \proofint_S \varphi(s,s',\tau) d\lambda(s') d(\mu(\tau))(s) \\
		&\quad - \proofint_S g(s) \proofint_S \varphi_n(s,s',\tau) d\lambda_n(s') d(\mu_n(\tau))(s),
	\\ 
		|\Delta_2(\tau)| &\le (L_1+L_3+M) d_\BL(\mu(\tau), \mu_n(\tau)) \\
		&\quad + \max\{L_2,M\} d_\BL(\lambda, \lambda_n).
	\end{align*}

	\paragraph{Initial states}

	As $g\in \lip_1(S), ~\|g\|_\infty \le 1$, we have
	\begin{equation*}
		\begin{multlined}
		\left|\proofint_S g(s) d(\mu(0))(s) - \proofint_S g(s) d(\mu_n(0))(s)\right| \\
		\le d_\BL(\mu(0), \mu_n(0)).
		\end{multlined}
	\end{equation*}

	\paragraph{Completing the proof}

	Given the above estimates,
	\begin{align*}
		& \left|\proofint_S gd\mu(t) - \proofint_S gd\mu_n(t)\right| \\
		& \quad \le \proofint_{[0,t]}(|\Delta_1(\tau)| + |\Delta_2(\tau)|)d\tau + d_\BL(\mu(0), \mu_n(0)) \\
		&\quad \le K \proofint_{[0,t]}(d_\BL(\mu(\tau), \mu_n(\tau)) + d_\BL(\lambda, \lambda_n))d\tau \\
		&\quad\quad + d_\BL(\mu(0), \mu_n(0)),
	\end{align*}
	where $K \coloneqq \max\{2L_3+3\max\{L_1,M\},3\max\{L_2,M\}\}$.
	Taking the supremum on the left-hand side, we find that
	\begin{align*}
		&d_\BL(\mu(t), \mu_n(t)) \\
		&\quad \le K \proofint_{[0,t]}\left(d_\BL(\mu(\tau), \mu_n(\tau)) + d_\BL(\lambda, \lambda_n)\right)d\tau \\
		&\quad\quad + d_\BL(\mu(0), \mu_n(0)).
	\end{align*}
	Applying Gr\"onwall's inequality and monotonicity of $e^{Kt}$,
	\begin{align*}
		&\sup_{t\in[0,T]} d_\BL(\mu(t), \mu_n(t)) \\
		&\quad \le - d_\BL(\lambda, \lambda_n) + (d_\BL(\lambda, \lambda_n) + d_\BL(\mu(0), \mu_n(0))) e^{KT}.
	\end{align*}
	Since $n\in\N$ is arbitrary and $\lim_{n\to\infty} d_\BL(\lambda, \lambda_n) = \lim_{n\to\infty} d_\BL(\mu(0), \mu_n(0)) = 0$ as both $\lambda_n \overset{n\to\infty}{\to} \lambda$ weakly and $\mu_n(0) \overset{n\to\infty}{\to} \mu(0)$ weakly, the conclusion \eqref{eq: approx} follows.
\end{proofE}

\begin{remark}
	\label{rem: approx}
	\autoref{thm: approx} gives a strong type of convergence: the approximation converges to the true dynamics uniformly on compact time intervals. This implies the pointwise convergence that $\mu_n(t) \overset{n\to\infty}{\to} \mu(t)$ weakly for all $t\in[0,\infty)$.
\end{remark}

\begin{remark}
	\label{rem: finite_approximations}
	Although \autoref{thm: approx} holds for general approximating measures $\lambda_n,\mu_n(0) \in \P(S)$---not only finite ones---it is most useful to consider the finite case, which guarantees that the computationally tractable system of ODEs \eqref{eq: n_ode} accurately reflects the true infinite-dimensional dynamics in finite time. We will see in \autoref{sec: long-time_analysis} that this does \emph{not} necessarily imply accuracy in the long-time limit.
\end{remark}

\begin{remark}
	\label{rem: fixed-in-time}
	\autoref{thm: approx} still holds for time-varying reference measures $\lambda$ and $\lambda_n$, under the additional assumptions that 1) $t\mapsto \lambda(t)$ and all $t\mapsto \lambda_n(t)$ are weakly continuous, and 2) $\lambda_n(t) \overset{n\to\infty}{\to} \lambda(t)$ weakly for all $t \in [0,\infty)$. The proof becomes more complicated, relying on a generalized version of Gr\"onwall's inequality \citep[Theorem~9]{dragomir2003some} and some additional continuity arguments, which we omit for conciseness.
\end{remark}

\begin{remark}
	\label{rem: empirical_distribution}
	Given a measure $\lambda\in\P(S)$, we may always construct an approximating measure $\lambda_n \coloneqq \frac{1}{n}\sum_{i=1}^n \delta_{s_i} \in\P(S)$ with $s_1,\dots,s_n \in S$ sampled independently and identically from $\lambda$, and in this case, the law of large numbers ensures that the weak convergence assumption in \autoref{thm: approx} holds $\lambda$-almost surely. The same goes for approximating $\mu(0)$.
\end{remark}

\subsection{Long-Time Analysis and Convergence to Equilibria}
\label{sec: long-time_analysis}

Although \autoref{thm: approx} shows that finite-dimensional approximations accurately reflect the evolutionary game's behavior over finite-time intervals under rather weak assumptions, it does \emph{not} automatically imply accurate approximations in the long-time limit. Specifically, \autoref{thm: approx} cannot be applied to assess the game's convergence to Nash equilibria using the finite approximations. To rigorously analyze such asymptotic behavior, we identify and introduce the key notion of ``choice paralysis:''

\begin{definition}
	\label{def: choice-mobile_trajectories}
	Consider a collection $\mathcal{X} = \{x_n: n\in\N\}$ of trajectories $x_n \colon [0,\infty) \to \Delta^{n-1}$ such that, for all $n\in\N$, there exists $\overline{x}_n \in\Delta^{n-1}$ such that $\lim_{t\to\infty} x_n(t) = \overline{x}_n$. The collection $\mathcal{X}$ is said to be \emph{choice-mobile} if
	\begin{equation}
		\lim_{t\to\infty}\sup_{n\in\N}\|x_n(t) - \overline{x}_n\|_1 = 0.
		\label{eq: choice-mobile}
	\end{equation}
	Otherwise, $\mathcal{X}$ is said to \emph{suffer from choice paralysis}.
\end{definition}

\hl{
\begin{remark}
	\label{rem: choice-mobile_EDMs}
	The definition of choice mobility can be naturally applied to collections of finite-dimensional EDMs (such as \eqref{eq: n_ode}). Specifically, a collection of finite-dimensional EDMs is choice-mobile if, for every collection of initial conditions $\{x_n(0) \in \Delta^{n-1} : n\in\mathbb{N}\}$, the resulting solutions $x_n\colon [0,\infty) \to \Delta^{n-1}$ to the EDMs form a choice-mobile collection of trajectories.
\end{remark}
}

\begin{remark}
	\label{rem: l1-norm}
	The $\ell_1$-norm in \eqref{eq: choice-mobile} cannot hastily be replaced with another norm, even if they induce the same topology on $\Rn$ for a fixed value of $n$, since the tightness of the bounds between equivalent norms on $\Rn$ in general depends on $n$. Using $\|\cdot\|_1$ induces a natural comparison between approximations of different dimensionality, as $\|x\|_1 = 1$ for all $x\in\Delta^{n-1}$, irrespective of the number of strategies $n\in\N$.
\end{remark}

We now consider what choice mobility represents in a game theoretic context. Mathematically, \eqref{eq: choice-mobile} corresponds to \emph{uniform} convergence $x_n(t) \overset{t\to\infty}{\to} \overline{x}_n$. The uniformity in $n$ requires every finite-dimensional approximation to converge to its associated equilibrium at some speed independent of the number of strategies available. This corresponds to finite-dimensional games that remain sufficiently ``mobile'' even as more and more strategies become available, i.e., giving the players more strategies to choose from does not significantly slow down the overall dynamics of the game.
Contrarily, if a population suffers from choice paralysis, progression towards equilibrium may slow down upon increasing the (finite) number of strategies; the more choices there are, the longer it takes players to figure out what they want to do. If the evolution rate is slowed down enough relative to the convergence rates toward equilibria at a fixed number of strategies, then convergence can no longer be guaranteed as the number of strategies increases towards infinity.
From this perspective, it should be expected that the choice mobility condition \eqref{eq: choice-mobile} is key in ensuring that the infinite-dimensional state $\mu(t)$ actually ``makes progress'' towards an equilibrium. We show that this is indeed the case in \autoref{thm: finite_approx}.

\begin{theoremE}[][end,text link=]
	\label{thm: finite_approx}
	Assume \autoref{ass: approx}.
	Let $\theta_n,x_n(0)\in\Delta^{n-1}$ for all $n\in\N$, and consider the approximation \eqref{eq: n_ode} with $\lambda_n \coloneqq \sum_{i=1}^n (\theta_n)_i \delta_{s_i}$ and initial condition $x_n(0)$. Assume that $\lambda$ and all $\lambda_n$ are fixed in time. If $\lambda_n \overset{n\to\infty}{\to} \lambda$ weakly and $\sum_{i=1}^n (x_n(0))_i \delta_{s_i} \overset{n\to\infty}{\to} \mu(0)$ weakly, then
	\begin{equation}
		\lim_{t\to\infty}\lim_{n\to\infty} d_\BL\Bigg(\sum_{i=1}^n (\overline{x}_n)_i \delta_{s_i} , \mu(t)\Bigg) = 0
		\label{eq: finite_approx-2}
	\end{equation}
	whenever $\mathcal{X} = \{x_n : n\in\N\}$ is choice-mobile with every $\lim_{t\to\infty} x_n(t) = \overline{x}_n$ for some collection $\{\overline{x}_n \in \Delta^{n-1} : n\in\N\}$. In this case,
	\begin{enumerate}
		\item $\mu(t) \overset{t\to\infty}{\to} \overline{\mu}$ weakly whenever $\sum_{i=1}^n (\overline{x}_n)_i \delta_{s_i} \overset{n\to\infty}{\to} \overline{\mu}$ weakly for some $\overline{\mu}\in\P(S)$, and
		\item $\sum_{i=1}^n (\overline{x}_n)_i \delta_{s_i} \overset{n\to\infty}{\to} \overline{\mu}$ weakly whenever $\mu(t) \overset{t\to\infty}{\to} \overline{\mu}$ weakly for some $\overline{\mu}\in\P(S)$.
	\end{enumerate}
\end{theoremE}

\begin{proofE}
%
	Suppose that $\lim_{t\to\infty}\sup_{n\in\N}\|x_n(t) - \overline{x}_n\|_1 = 0$ for some collection $\{\overline{x}_n\in\Delta^{n-1} : n\in\N\}$. Let $T\in(0,\infty)$. By the triangle inequality and \autoref{thm: approx},
	\begin{align*}
		&\limsup_{n \to \infty} d_\BL\left(\proofsum_{i=1}^n (\overline{x}_n)_i \delta_{s_i} , \mu(T)\right) \\ 
		&\quad \le \limsup_{n \to \infty} d_\BL\left(\proofsum_{i=1}^n (\overline{x}_n)_i \delta_{s_i}, \proofsum_{i=1}^n (x_n(T))_i\delta_{s_i}\right) \\
		&\quad\quad + \limsup_{n\to\infty} \sup_{\tau\in[0,T]} d_\BL \left(\proofsum_{i=1}^n (x_n(\tau))_i\delta_{s_i} , \mu(\tau)\right) \\
		&\quad \le \sup_{n \in \N} d_\BL\left(\proofsum_{i=1}^n (\overline{x}_n)_i \delta_{s_i}, \proofsum_{i=1}^n (x_n(T))_i\delta_{s_i}\right).
	\end{align*}
	Since $T\in(0,\infty)$ is arbitrary, this gives that
	\begin{align*}
		&\limsup_{T\to\infty} \limsup_{n\to\infty} d_\BL\left(\proofsum_{i=1}^n (\overline{x}_n)_i \delta_{s_i} , \mu(T)\right) \\
		&\quad \le \limsup_{T\to\infty}\sup_{n\in \N} \sup_{\substack{g\in\lip_1(S) \\ \|g\|_\infty \le 1}} \left| \proofsum_{i=1}^n ((\overline{x}_n)_i - (x_n(T))_i) g(s_i) \right| \\
		&\quad \le \limsup_{T\to\infty} \sup_{n\in\N} \proofsum_{i=1}^n \left|(\overline{x}_n)_i - (x_n(T))_i\right| = 0.
	\end{align*}
	Hence, $\lim_{t\to\infty} \lim_{n\to\infty} d_\BL\left( \proofsum_{i=1}^n (\overline{x}_n)_i \delta_{s_i} , \mu(t)\right) = 0$, which proves \eqref{eq: finite_approx-2}. The two final enumerated results follow immediately from the triangle inequality.
\end{proofE}

\begin{remark}
	\label{rem: fixed-in-time-2}
	As was the case for \autoref{thm: approx}, \autoref{thm: finite_approx} also still holds for time-varying reference measures $\lambda$ and $\lambda_n$, under the same additional assumptions outlined in \autoref{rem: fixed-in-time}. The proof of \autoref{thm: finite_approx} is essentially unchanged in this case.
\end{remark}

Let us dissect \autoref{thm: finite_approx}.
The result \eqref{eq: finite_approx-2} ensures that, as the resolution of the approximation \eqref{eq: n_ode} becomes finer, the approximation accurately reflects the long-time behavior of the true game. \hl{Specifically, as time increases, $\mu(t)$ ``tracks'' the trajectory of the finite distribution $\sum_{i=1}^n (\overline{x}_n)_i \delta_{s_i}$ as the resolution $n$ increases.} The two final enumerated results ensure \hl{exact} coincidence between an asymptotically stable equilibrium $\overline{\mu}$ of the infinite-dimensional game and the limit of the asymptotically stable equilibria $\overline{x}_n$ of the approximations. In particular, they assert that, when the discrete distribution generated by $\overline{x}_n$ converges towards some limiting distribution $\overline{\mu}$, then $\mu(t)$ must approach this same distribution, and, on the other hand, when $\mu(t)$ approaches some limiting distribution $\overline{\mu}$, the discrete distribution generated by $\overline{x}_n$ must converge towards the same distribution. \hl{In other words, choice mobility ensures that the infinite-strategy dynamics converge if and only if the finite approximations converge, and they must converge to the same limiting distribution.\footnote{\hl{This amounts to being able to interchange the order of the limits in \eqref{eq: finite_approx-2}.}}} This shows that we may ``trust'' finite approximations \hl{of sufficiently fine resolution to accurately reflect the short- and long-time behavior} of infinite-strategy games when the approximations satisfy choice mobility. \hl{We remark that the convergence results in \autoref{thm: approx} and \autoref{thm: finite_approx} are asymptotic in $n$. Developing non-asymptotic rates of convergence to explicitly quantify the relationship between the approximation's resolution and its resulting accuracy poses an interesting direction for future work.}

\subsection{Emergence of Choice Paralysis}

\hl{We now identify one way that choice paralysis emerges, namely, if the rate of strategy switching decreases towards zero as the number of strategies increases.}

\begin{propositionE}[][end,text link=]
	\label{prop: choice_paralysis-decaying_switch_rate}
	Let $\theta_n,x_n(0) \in \Delta^{n-1}$ for all $n\in\N$, and consider the approximation \eqref{eq: n_ode} with $\lambda_n \coloneqq \sum_{i=1}^n (\theta_n)_i \delta_{s_i}$ and initial condition $x_n(0)$. Suppose for all $n\in\N$ that $\lim_{t\to\infty}x_n(t) = \overline{x}_n$ for some $\overline{x}_n\in \Delta^{n-1}$. If there exist $\epsilon > 0$ and a sequence $\{M_n \in \Rpl : n\in\N\}$ such that $M_n \overset{n\to\infty}{\to} 0$ and, for all $n\in\N$, it holds that $\|x_n(0) - \overline{x}_n\|_1 > \epsilon$ and
	\begin{equation}
		\|\dot{x}_n(t)\|_1 \le M_n ~ \text{for all $t\in [0,\infty)$},
		\label{eq: velocity_decay}
	\end{equation}
	then $\mathcal{X} = \{x_n \colon n\in\N\}$ suffers from choice paralysis. In particular, \eqref{eq: velocity_decay} holds when the revision protocol $\hat{\varphi}_n$ satisfies $|\hat{\varphi}_n(i,j,\theta)| \le \frac{1}{2}M_n$ for all $i,j\in\{1,\dots,n\}$, $\theta \in \Delta^{n-1}$.
\end{propositionE}

\begin{proofE}
	Let $t\in[0,\infty)$. For all $n\in\N$,
	\begin{align*}
		\|x_n(t) - \overline{x}_n\|_1 &= \left\| \proofint_{[0,t]}\dot{x}_n(\tau) d\tau + x_n(0) - \overline{x}_n\right\|_1 \\
		&\ge \|x_n(0) - \overline{x}_n\|_1 - \proofint_{[0,t]}\|\dot{x}_n(\tau)\|_1 d\tau,
	\end{align*}
	Therefore,
	\begin{align*}
		\limsup_{n\to\infty}\|x_n(0) - \overline{x}_n\|_1 
		&\le \limsup_{n\to\infty} \|x_n(t) - \overline{x}_n\|_1 \\
		&\quad + \limsup_{n\to\infty}\proofint_{[0,t]}\|\dot{x}_n(\tau)\|_1d\tau,
	\end{align*}
	implying that
	\begin{equation*}
	\begin{aligned}
		\sup_{n\in\N}\|x_n(t) - \overline{x}_n\|_1 
		&\ge \limsup_{n\to\infty}\|x_n(0) - \overline{x}_n\|_1 \\
		&\quad - \limsup_{n\to\infty} \proofint_{[0,t]}\|\dot{x}_n(\tau)\|_1 d\tau \\
		&\ge \epsilon - \limsup_{n\to\infty} M_n t = \epsilon,
	\end{aligned}
	\end{equation*}
	so $\mathcal{X} = \{x_n : n\in\N\}$ suffers from choice paralysis.

	If $|\hat{\varphi}_n(i,j,\theta)| \le \prooffrac{1}{2}M_n$ for all $i,j$ and all $\theta$, then \eqref{eq: n_ode} gives that
	\begin{align*}
		|(\dot{x}_n(t))_i| &\le \prooffrac{1}{2} M_n \proofsum_{j=1}^n \left(\lambda_n(s_i) (x_n(t))_j + (x_n(t))_i \lambda_n(s_j)\right) \\
		&= \prooffrac{1}{2} M_n (\lambda_n(s_i) + (x_n(t))_i)
	\end{align*}
	for all $n\in\N$ and all $i\in\{1,\dots,n\}$, and therefore
	\begin{equation*}
		\|\dot{x}_n(t)\|_1 
		\le \prooffrac{1}{2} M_n \proofsum_{i=1}^n(\lambda_n(s_i) + (x_n(t))_i) = M_n,
	\end{equation*}
	so \eqref{eq: velocity_decay} holds since $t\in [0,\infty)$ is arbitrary.
\end{proofE}

\autoref{prop: choice_paralysis-decaying_switch_rate} is not restricted to fixed reference measures; it holds for time-varying $\lambda_n$, e.g., in the replicator dynamics.

\section{Example}
\label{sec: example}

We now construct an explicit example illustrating choice paralysis and inconsistency between the infinite-strategy dynamics and the finite approximations. Consider the game on $S = [0,1]$ defined by $F(\mu)(s) = \int_S f(s,s') d\mu(s')$ with
\begin{equation}
	f(s,s') = -1 ~ \text{if $s=s'$}, \quad f(s,s') = 0 ~ \text{otherwise},
	\label{eq: discts_payoff}
\end{equation}
evolving according to the replicator dynamics (recall \autoref{ex: revision_protocols}). This game penalizes players that choose the same strategy as another player, and hence is often referred to as the ``anticoordination game'' \citep{sandholm2010population}. It is straightforward to see that $\dot{\mu}(t)(B) = \int_B \left( F(\mu(t))(s) - \int_S F(\mu(t))(s') d(\mu(t))(s') \right) d(\mu(t))(s) = 0$ for every set $B\in\B(S)$, whenever $\mu(t)$ is absolutely continuous with respect to Lebesgue measure. In particular, this shows that the infinite-strategy evolutionary game gets completely stuck at such population states; $\mu(t) = \mu(0)$ for all $t\in [0,\infty)$, for every $\mu(0)$ with (Lebesgue) density.

On the other hand, the approximation \eqref{eq: n_ode} reduces to
\begin{equation}
	(\dot{x}_n(t))_i = -(x_n(t))_i^2 + (x_n(t))_i \displaystyle \sum_{j=1}^n (x_n(t))_j^2
	\label{eq: approx_example}
\end{equation}
for all $i\in\{1,\dots,n\}$ for every $n\in\N$, which corresponds to the finite-dimensional game defined by $\hat{F}_n(\theta) = - \theta$. It is straightforward to show that $\overline{x}_n = \frac{1}{n}\mathbf{1}_n$ is the only stationary point of these dynamics in the relative interior of $\Delta^{n-1}$, where $\mathbf{1}_n$ denotes the $n$-vector of all ones, and that $\overline{x}_n$ is also this (strictly concave) game's unique Nash equilibrium \citep[Corollary~3.1.4]{sandholm2010population}. As such, the stationary point $\overline{x}_n = \frac{1}{n}\mathbf{1}_n$ is globally asymptotically stable over the relative interior of $\Delta^{n-1}$ \citep[Theorem~7.2.4]{sandholm2010population} for every $n\in\N$%
,
suggesting that the game converges towards the uniform distribution on $S$ when initialized with full support
. Therefore, the approximations fail to accurately model the game's behavior in the long-time limit. Indeed, the dynamics suffer from choice paralysis. Intuitively, this is because, in the vicinity of the stationary point, $\|\dot{x}_n(t)\|_1 \le 2 \|x_n(t)\|_2^2 \approx 2\|\overline{x}_n\|_2^2 = \frac{2}{n} \overset{n\to \infty}{\to} 0$, so paralysis emerges (recall \autoref{prop: choice_paralysis-decaying_switch_rate}). This can occur even at states that are uniformly bounded away from $\overline{x}_n$ in $\ell_1$. For instance, let $\epsilon > 0$, and for all $n\in\N$ define $m_n\in\N$ to be the ceiling of $\frac{\epsilon n}{2}$ and consider $x_n(0) = \big((\frac{1}{n} + \frac{\epsilon}{2m_n})\mathbf{1}_{m_n}, (\frac{1}{n} - \frac{\epsilon}{2 m_n}) \mathbf{1}_{m_n}, \frac{1}{n} \mathbf{1}_{n-2 m_n}\big)$. It holds that $x_n(0) \in \Delta^{n-1}$, that $\|x_n(0) - \overline{x}_n\|_1 = \epsilon$, and that $\|\dot{x}_n(0)\|_1 \le 2 \|x_n(0)\|_2^2 = \frac{2}{n} + \frac{\epsilon^2}{m_n} \overset{n\to \infty}{\to} 0$, so the collection of approximations exhibits paralysis immediately at $t=0$ despite the fact that every individual approximation converges to its associated equilibrium.

\hl{The above example can be extended to one that satisfies Assumption 1 (via \autoref{prop: bnn_pcd_approx}) by replacing the payoff function by $f(s,s') = -h(|s-s'|)$, where $h \colon [0,1]\to[0,1]$ is a smooth bump function with compact support and a unique maximizer at $0$ giving $h(0) = 1$. In this case, the right-hand side of the approximation \eqref{eq: approx_example} becomes linear (and hence Lipschitz and smooth) in the values $h(|s_i - s_j|)$. Therefore, the bump function $h$ can be chosen close enough to the discontinuous payoff function \eqref{eq: discts_payoff} such that $\|\dot{x}_n(0)\|_1$ from above still decreases at a rate of $\frac{1}{n}$ as $n\to \infty$, and hence choice paralysis still arises in light of \autoref{prop: choice_paralysis-decaying_switch_rate}.}

\appendix

\section{Proofs}
\label{sec: proofs}
\printProofs

\section*{Acknowledgments}
The author thanks Murat Arcak for their constructive feedback and insightful discussions.

\bibliographystyle{abbrvunsrtnat}
\bibliography{tex/references.bib}


\end{document}